\documentclass[twocolumn]{aastex631}
\usepackage{graphics,epsf}
\usepackage{amsmath}                
\usepackage{amsfonts}               
\usepackage{amssymb}                
\usepackage{epsfig}                 
\usepackage{appendix}
\usepackage{graphicx}
\usepackage{float}
\usepackage{color}
\usepackage{multirow}
\usepackage{colortbl}
\usepackage[para,online,flushleft]{threeparttable}

\hypersetup{citecolor=blue, 
            linkcolor=red, 
            menucolor=blue, 
            urlcolor=blue}  

\newcommand{\m}{{~\rm m}}
\newcommand{\km}{{~\rm km}}
\newcommand{\s}{{~\rm s}}
\newcommand{\h}{{~\rm h}}

\newcommand{\g}{{~\rm g}}
\newcommand{\G}{{~\rm G}}
\newcommand{\K}{{~\rm K}}

\begin{document}

\title{The jets and the neutron star kick velocity of the supernova remnant CTB 1} 


\author[0000-0002-3592-1526]{Ealeal Bear}
\affiliation{Department of Physics, Technion, Haifa, 3200003, Israel;  ealeal44@technion.ac.il; soker@physics.technion.ac.il}

\author[0000-0003-0375-8987]{Noam Soker}
\affiliation{Department of Physics, Technion, Haifa, 3200003, Israel;  ealeal44@technion.ac.il; soker@physics.technion.ac.il}

\begin{abstract}
We identify jet-shaped morphology in the core-collapse supernova remnant (SNR) CTB 1 that includes two opposite structural features. We identify these as the imprints of a pair of jets that were among the last jets to explode the massive stellar progenitor of CTB 1. We find the projected angle between the jets’ axis and the direction of the pulsar velocity, which is the neutron star natal kick, to be $\alpha=78^\circ$. We tentatively identify possible signatures of a second pair of opposite jets along a different direction. If this identification holds, SNR CTB 1 has a point-symmetric structure. The morphology and large angle of the jets’ axis to kick velocity are expected in the jittering jets explosion mechanism (JJEM) of core-collapse supernovae. 
\end{abstract}


\section{Introduction}
\label{sec:intro}

Neutron stars (NSs) are born in core-collapse supernova (CCSN) explosions with a typical natal kick velocity of $v_{\rm NS} \simeq 200 - 500\km \s^{-1}$ (e.g.,  \citealt{Kapiletal2022}). In the frame of the jittering jets explosion mechanism (JJEM), an important parameter is the angle $\alpha$ between the natal kick direction and the axis of the two opposite jets that shape the ejecta the most. The symmetry axis might be determined by two opposite ears (two opposite protrusions from the main ejecta) or by the axis of a barrel-shaped supernova remnant (SNR). Earlier studies (\citealt{BearSoker2018kick, Soker2022SNR0540}) of the observed cumulative distribution function of the angle $\alpha$, ${\rm W}_{\alpha}({\rm obs})$, found that the angle $\alpha$ tends to be large. Namely, the kick velocity avoids small angles relative to the main jets' axis. In the JJEM there are several to a few tens of jet-launching episodes, hence the same number of jets' axes. By the main jets' axis, we refer to the pair of opposite jets that inflate the largest ears or lobes.

The JJEM can account for this avoidance of small angles $\alpha$ (e.g., \citealt{Soker2023kick}) if the kick velocity is due to the gravitational tug-boat mechanism. In the tug-boat mechanism (e.g., \citealt{Wongwathanaratetal2013}) a clump, or several clumps, that are ejected by the explosion mechanism, gravitationally pulls the NS and accelerates it to acquire its natal kick velocity. In the JJEM the final jets, which mostly shape the outer ejecta, prevent the formation of dense clumps along their propagation direction, hence no acceleration of the NS along the jets' axis. Else, the clumps that feed the final accretion disk and the clumps that accelerate the NS, come from the same region. Therefore, the acceleration is in the plane of the last accretion disk and hence tends to be at a large angle to the axis of the last jets.

\section{The morphology of SNR CTB~1}
\label{sec:point}

\begin{figure*}[!t]
\includegraphics[trim=0.0cm 11.0cm 0.0cm 0.0cm ,clip, scale=0.90]{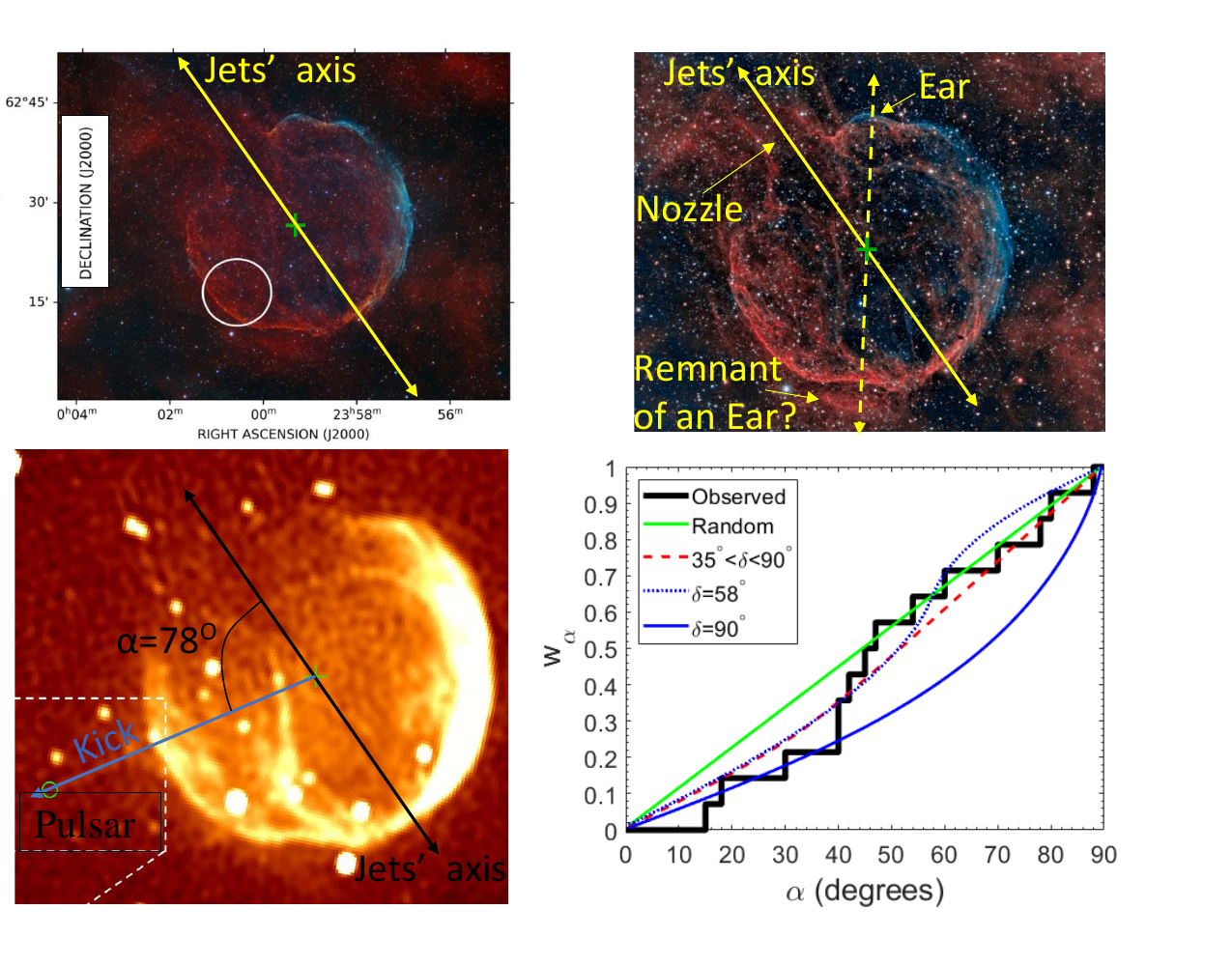}
\caption{Upper left panel: An image adapted from \cite{ReyesIturbideetal2024} based on Tomasz Zwolinski (http://astrozdjecia.pl/); white circle from their image.  Red and blue are for H$\alpha$ and [O III]. We added the double-headed arrow to signify the main jets' axis.  
Upper right panel: Our identification of morphological features on a picture adapted from Kimberly Sibbald (https://www.flickr.com/photos/spacepaparazzi/51748096139/in/dateposted-public/). 
Lower left panel: A radio image from \cite{Schinzeletal2019}. We added the jets' axis and the kick velocity.  
Lower right panel: Observed cumulative distribution function (black step-function)
and four different theoretical cumulative distribution functions. 
}
\label{fig:distribution}
\end{figure*}

We take the image in H$\alpha$ and [O III] from \cite{ReyesIturbideetal2024} (upper-left panel of Figure \ref{fig:distribution}). We assume that the `nozzle' on the north side was shaped by one of the last jets out of many that exploded the star (e.g., \citealt{BearSoker2018kick} for other SNRs).  On the opposite side of the center (green plus) there is a bright bar/filament, with a short thin outer filament. We assume that this bright bar/filament structure is the rim of an opening (nozzle) through which the counter jet expanded to the ambient gas. Unequal opposite ears/lobes/nozzles in CCSN remnants is very common and has an explanation in the JJEM \citep{Soker2023counterjet}. The line that we draw from the nozzle in the north to the bar in the south is the jets' axis of this SNR.   

The old SNR CTB~1, age of $47.60 \pm 0.80 ~{\rm kyr}$ \citep{Bruzewskietal2023}, was heavily influenced by the ambient gas, circumstellar and interstellar, i.e., the SNR was substantially decelerated such that the pulsar (green circle in the lower-left panel) is already outside the SNR. We, nonetheless, speculate on another pair of jets that left an imprint on SNR CTB~1. We mark the tentatively assumed axis of this pair of jets with a dashed-double-headed yellow arrow on the upper-right panel. We identify this axis by the ear in the north and a possible remnant of an ear in the south. If this second pair of jets is real, then this SNR has a point-symmetric morphology, which most likely is due to shaping by jets, rather than by ejecta interaction with ambient gas (e.g., \citealt{Soker2024Rev}). 

\section{Kick velocity}
\label{sec:angle}

The projected (on the plane of the sky) angle between the jets' main axis and the kick velocity of the pulsar ($\simeq 500 \km \s^{-1}$; \citealt{Bruzewskietal2023}) is $\alpha=78^\circ$ (lower-left panel).  

We add this new angle $\alpha=78^\circ$ to the previous 13 cases, 12 from \cite{BearSoker2018kick} and one from \cite{Soker2022SNR0540}, to build an updated observed cumulative distribution function ${\rm W}_{\alpha}({\rm obs})$; the thick-black step function in the lower-right panel of Figure \ref{fig:distribution}. 
We also draw four theoretical cumulative distribution functions according to \cite{Soker2023kick} by the behavior of the three-dimensional (3D) angle between the jets' axis and the kick velocity $\delta$: One completely random in 3D (solid-green), one limited-random in which the angle in 3D is random in the range of $\delta_c=35^\circ < \delta \le 90^\circ$ and zero for $\delta \le \delta_c = 35^\circ $ (dashed-red line), one for which the 3D angle is fixed at $\delta=\delta_F=58^\circ$, and one for which the 3D angle is fixed at $\delta=\delta_F=90^\circ$ (kick velocity always perpendicular to jets' axis in 3D).  
\cite{Soker2023kick} analyzed 13 SNRs and found the best fits to be either the limited-random distribution with $\delta_c=30^\circ$ or the fixed-angle with $\delta_F=55^\circ$. With the 14 SNRs, we find the best fittings to have larger angles of $\delta_c=35^\circ$ and $\delta_F=58^\circ$. This strengthens our earlier claim  \citep{BearSoker2018kick} that the jet-kick angle avoids small angles. 

\section{Summary}
\label{sec:Summary}

Motivated by the new papers \cite{Bruzewskietal2023} and \cite{ReyesIturbideetal2024} we examined the CCSN remnant CTB~1. We identified a main jets' axis. We further speculated on another axis of two jets (dashed-yellow-double-headed arrow) to tentatively identify a point-symmetric morphology. A point-symmetric morphology is a unique prediction of the JJEM. The identification of point-symmetric morphologies in three CCSN remnants in 2023 (N63A, Vela, SN 1987A) might be a breakthrough in establishing the JJEM as the main explosion mechanism of CCSNe \citep{Soker2024Rev}. 

We find the projected angle on the plane of the sky between the kick velocity and the jets' axis of CTB~1 to be $\alpha=78^\circ$. This is also compatible with the expectation of the JJEM. 



\end{document}